\documentclass[prd,preprint,aps,amsfonts,amsmath,amssymb]{revtex4}
\allowdisplaybreaks
\begin{document}
\title{Gravitational-Wave Inspiral of Compact Binary Systems\\ to 7/2
Post-Newtonian Order} \author{Luc Blanchet$^{1,3}$, Guillaume
Faye$^4$, Bala R. Iyer$^2$ and Benoit Joguet$^3$} \affiliation{$^1$
D\'epartement d'Astrophysique Relativiste et de Cosmologie (UMR 8629
du CNRS),\\ Observatoire de Paris, 92195 Meudon Cedex, France}
\affiliation{$^2$ Raman Research Institute, Bangalore 560 080, India}
\affiliation{$^3$ Institut d'Astrophysique de Paris, 98bis boulevard
Arago, Paris, France} \affiliation{$^4$ Theoretisch-Physikalisches
Institut, Friedrich-Schiller-Universit\"at,\\ Max-Wien-Pl. 1, 07743
Jena, Germany} \date{\today}
\widetext
\begin{abstract}
The inspiral of compact binaries, driven by gravitational-radiation
reaction, is investigated through 7/2 post-Newtonian (3.5PN) order
beyond the quadrupole radiation. We outline the derivation of the
3.5PN-accurate binary's center-of-mass energy and emitted
gravitational flux. The analysis consistently includes the
relativistic effects in the binary's equations of motion and multipole
moments, as well as the contributions of tails, and tails of tails, in
the wave zone. However the result is not fully determined because of
some physical incompleteness, present at the 3PN order, of the model
of point-particle and the associated Hadamard-type self-field
regularization. The orbital phase, whose prior knowledge is crucial
for searching and analyzing the inspiral signal, is computed from the
standard energy balance argument.
\end{abstract}


\maketitle


A few years ago it was recognized \cite{3mn} that improved waveform
modelling is crucial to construct templates for searching and
measuring gravitational-waves from inspiralling compact binaries with
laser-interferometric detectors like LIGO and VIRGO. Since a large
number of orbital cycles are observable in the frequency band of the
detectors, the measurement, using the technique of matched-filtering,
will be extremely sensitive to those parameters that affect the
inspiral rate and thus the orbital phase evolution. The orbital phase
(which is driven by gravitational-radiation reaction) is therefore the
crucial quantity to be monitored for these
experiments. Measurement-accuracy analyses \cite{CFPS} have shown that
a very high post-Newtonian prediction, probably the third
post-Newtonian, or even the 3.5PN one (i.e. $1/c^7$), in the case of
neutron-star binaries, would be required.  Only then the templates
would be accurate enough over most of the inspiral phase, with reduced
cumulative phase lags, so that the phasing errors are not significant
when one attempts to extract the values of the binary's parameters
(essentially the masses and spins) from the data. Having in hand such
high-order post-Newtonian expressions, one could apply resummation 
methods to further improve the convergence of the
post-Newtonian series, and make it even more accurate for searches as
well as parameter estimations \cite{DIS,BD}. In this Letter, to provide
the essential theoretical input for gravitational-wave data analysis
\cite{3mn,CFPS,DIS,BD}, we compute the orbital phase of compact binaries, 
both in the time and frequency domains, in the adiabatic 
approximation, at the 3.5PN order. 
Numerical relativity or approaches such
as \cite{BD} could describe the plunge and merger phases. The latter
approach starts from the post-Newtonian expansion and goes beyond the 
adiabatic approximation.
Appropriate to the inspiral regime \cite{3mn}, we treat the
compact bodies as structureless, non-spinning point-particles, moving
on quasi-circular orbits. Spin effects are known up to 2.5PN order
\cite{KWW} and may be added if necessary.

The first ingredient in the theoretical analysis is the equation of
motion of the binary, which is used primarily for the calculation of
the center-of-mass energy $E$ that is conserved in the absence of
gravitational-radiation reaction. Recently, the equation of motion of
compact binaries at the 3PN order has been obtained by means of two
different methods, with equivalent results. Jaranowski and Sch\"afer
\cite{JaraS}, and Damour, Jaranowski and Sch\"afer \cite{DJS,DJSb},
employ the Arnowitt-Deser-Misner Hamiltonian formalism of general
relativity, while Blanchet and Faye \cite{BF,BFreg}, and Andrade,
Blanchet and Faye \cite{ABF}, proceed with the post-Newtonian
iteration of the Einstein field equations in harmonic
coordinates. Since the binary's orbit would have been circularized by
radiation reaction, the equation of motion is of the form
\begin{equation}\label{1}
{d v^i\over dt} = - \omega^2 x^i + \frac{1}{c^5} F_{\rm reac}^i\;,
\end{equation}
where $x^i=y_1^i-y_2^i$ is the vector separation between the two
particles, $v^i=d x^i/dt$ the relative velocity, and $\omega$ the
orbital angular frequency ($\omega=2\pi/P$, where $P$ is the
period). We denote by $F_{\rm reac}^i$ the standard radiation reaction
--- a resistive force opposite to the relative velocity, which arises
dominantly at the 2.5PN order. Through 3PN order, the orbital
frequency is related to the distance $r=|{\bf x}|$ in harmonic
coordinates ({\it via} the post-Newtonian parameter
$\gamma=\frac{Gm}{rc^2}$) by \cite{BF}
\begin{eqnarray}\label{2}
 \omega^2 &=& {G m\over r^3}\biggl\{ 1+\left(-3+\nu\right) \gamma +
 \left(6+\frac{41}{4}\nu +\nu^2\right) \gamma^2 \nonumber\\
 &+&\left(-10+\left[-\frac{67759}{840}+\frac{41}{64}\pi^2
 +22\ln\left(\frac{r}{r'_0}\right)+\frac{44}{3}\lambda \right]\nu
 +\frac{19}{2}\nu^2+\nu^3\right) \gamma^3 \biggr\}\;.
\end{eqnarray}
Mass parameters are the total mass $m=m_1+m_2$ and the symmetric mass
ratio $\nu=m_1m_2/m^2$ satisfying $0<\nu\leq 1/4$ (the reduced mass is
then $\mu=m\nu$).  The 3PN coefficient depends on two arbitrary
constants: a length scale $r_0'$ entering the logarithm, and the
constant $\lambda$. It was shown in Ref. \cite{BF} that $r_0'$ is
merely linked with the choice of harmonic coordinates, and has
therefore no physical meaning, as it can be eliminated by a change of
gauge. By contrast, $\lambda$ represents a physical indeterminacy, in
the form of a purely numerical constant (e.g. a rational fraction),
and is probably associated with an incompleteness of the Hadamard-type
method for regularizing the infinite self-field of point-particles
\cite{BFreg}, which is used to cope with the model of compact objects
idealized by Dirac functions (for general, non-circular orbits, it is
impossible to re-absorb $\lambda$ into a redefinition of the gauge
constant $r_0'$). The presence of $\lambda$ may be associated with the 
fact that many integrals composing the equation of motion, when taken 
{\it individually}, start depending, from the 3PN order, on the internal
structure of the bodies, even in the limit where their size tends to
zero. However, when considering the {\it full} equation of motion, 
we finally expect $\lambda$ to be 
independent of the internal structure of the compact bodies.
The constant $\lambda$ is equivalent to the static ambiguity
parameter $\omega_{\rm static}$ introduced in Refs. \cite{JaraS,DJS},
in the sense that $\lambda=-\frac{3}{11}\omega_{\rm
static}-\frac{1987}{3080}$. Recently, the value $\omega_{\rm
static}=0$ has been obtained by means of a dimensional regularization
suplementing the ADM-Hamiltonian formalism \cite{DJSb}. This result
would mean that $\lambda = -\frac{1987}{3080}$ (but we keep $\lambda$
unspecified in this discussion).

{}From now on we shall use in place of the angular frequency $\omega$
the dimensionless variable $x=\left(\frac{\pi G m
f}{c^3}\right)^{2/3}$, where $f=2/P=\omega/\pi$ is the frequency of
the gravitational-wave signal at the dominant harmonic. By inverting
Eq. (\ref{2}) one finds $\gamma$ in terms of the variable $x$, which
we shall now consider as an alternative ordering post-Newtonian
parameter,
\begin{eqnarray}\label{3}
 \gamma &=& x \biggl\{ 1+\left(1-\frac{1}{3}\nu\right) x +
 \left(1-\frac{65}{12}\nu\right) x^2 \nonumber\\
 &+&\left(1+\left[-\frac{10151}{2520}-\frac{41}{192}\pi^2
 -\frac{22}{3}\ln\left(\frac{r}{r'_0}\right)-\frac{44}{9}\lambda
 \right]\nu +\frac{229}{36}\nu^2+\frac{1}{81}\nu^3\right) x^3 \biggr\}\;.
\end{eqnarray}
As the 3PN equation of motion for general orbits derives from a
Lagrangian \cite{ABF} (neglecting the radiation reaction), one can
straightforwardly compute the associated 3PN conserved energy. The
result, when specialized to circular orbits, reads
\begin{eqnarray}\label{4}
 E &=& -{\mu c^2 \gamma\over 2} \biggl\{ 1
 +\left(-\frac{7}{4}+\frac{1}{4}\nu\right) \gamma +
 \left(-\frac{7}{8}+\frac{49}{8}\nu +\frac{1}{8}\nu^2\right) \gamma^2
 \nonumber\\
 &+&\left(-\frac{235}{64}+\left[\frac{106301}{6720}-\frac{123}{64}\pi^2
 +\frac{22}{3}\ln \left(\frac{r}{r_0'}\right)
 -\frac{22}{3}\lambda\right]\nu+\frac{27}{32}\nu^2
 +\frac{5}{64}\nu^3\right)\gamma^3 \biggr\}\;.
\end{eqnarray}
The good thing to do next is to re-express this energy in terms of the
post-Newtonian parameter $x$. Indeed, as $x$ is directly related to
the orbital period, the energy will be form invariant (the same in
different coordinate systems). We find \cite{DJS,BF}
\begin{eqnarray}\label{5}
 E &=& -{\mu c^2 x\over 2} \biggl\{ 1
 +\left(-\frac{3}{4}-\frac{1}{12}\nu\right) x +
 \left(-\frac{27}{8}+\frac{19}{8}\nu -\frac{1}{24}\nu^2\right) x^2
 \nonumber\\
 &+&\left(-\frac{675}{64}+\left[\frac{209323}{4032}-\frac{205}{96}\pi^2
 -\frac{110}{9}\lambda\right]\nu-\frac{155}{96}\nu^2
 -\frac{35}{5184}\nu^3\right)x^3 \biggr\}\;.
\end{eqnarray}
As expected the latter expression is free of the unphysical gauge
constant $r_0'$. Since it can be checked that for circular orbits
there are no terms of order $x^{7/2}$, the energy (\ref{5}) is in fact
valid up to the 3.5PN order. In the test-mass limit $\nu\to 0$, we
recover the energy of a particle with mass $\mu$ in a Schwarzschild
background of mass $m$, i.e. $E_{\rm test}=\mu
c^2\left[(1-2x)(1-3x)^{-1/2}-1\right]$, when developed to the 3.5PN
order.

The second ingredient in this analysis concerns the gravitational
wave-form generated by the compact binary. More precisely, we need to
compute the binary's total energy flux at infinity, or gravitational
luminosity ${\cal L}$, in the post-Newtonian approximation. This
calculation should take into account the relativistic corrections
linked with the description of the source (multipole moments), as well
as the non-linear effects in the propagation of the waves from the
source to the far zone. We have applied here a particular
wave-generation formalism \cite{BDI,B98tail,B98mult}, valid for
slowly-moving sources, in which the exterior field is parametrized by
some specific multipole moments, formally valid to any post-Newtonian
order \cite{B98mult}, and where the observables at infinity are
connected to the source moments by some non-linear (post-Minkowskian)
functional relations, taking into account the various effects of tails
(see e.g. \cite{B98tail}). The formalism has already been specialized
to the case of inspiral wave-forms at the 2.5PN level by Blanchet,
Damour and Iyer \cite{BDI95}. Furthermore, a different formalism,
devised by Will and Wiseman \cite{WWi96}, was independently applied to
this problem and reached equivalent results, reported jointly in
Ref. \cite{BDIWW95}, at the 2PN order. The crucial input of any
post-Newtonian computation of the flux is the mass quadrupole moment
(indeed the required post-Newtonian precision of the higher moments is
smaller). The 3PN quadrupole moment for circular binary orbits, say
$I_{ij}=\mu\left(A~\!{\hat x}_{ij}+B~\!\frac{r^2}{c^2}~\!{\hat
v}_{ij}\right)$, where we neglect a 2.5PN term and denote e.g. ${\hat
x}_{ij}=x_ix_j-\frac{1}{3}\delta_{ij}r^2$, has been obtained recently
by Blanchet, Iyer and Joguet \cite{BIJ01}, who find the result
\begin{subequations}\label{6}\begin{eqnarray}
 A &=& 1+\left(-\frac{1}{42}-\frac{13}{14}\nu\right) \gamma +
 \left(-\frac{461}{1512}-\frac{18395}{1512}\nu
 -\frac{241}{1512}\nu^2\right) \gamma^2 \nonumber\\
 &+&\left(\frac{395899}{13200}-\frac{428}{105}\ln\left(\frac{r}{r_0}\right)
 +\left[\frac{139675}{33264}-\frac{44}{3}\ln\left(\frac{r}{r'_0}\right)
 -\frac{44}{3}\xi-\frac{88}{3}\kappa\right]\nu \right.\nonumber\\
 &+&\left.\frac{162539}{16632}\nu^2+\frac{2351}{33264}\nu^3\right)
 \gamma^3\;, \\ B &=& \frac{11}{21}-\frac{11}{7}\nu +
 \left(\frac{1607}{378}-\frac{1681}{378}\nu+\frac{229}{378}\nu^2\right)
 \gamma \nonumber\\
 &+&\left(-\frac{357761}{19800}+\frac{428}{105}\ln\left(\frac{r}{r_0}\right)
 +\left[-\frac{75091}{5544}+\frac{44}{3}\zeta\right]\nu
 +\frac{35759}{924}\nu^2+\frac{457}{5544}\nu^3\right) \gamma^2\;.
\end{eqnarray}\end{subequations}$\!\!$
Note the two types of logarithms entering these formulas at the 3PN
order. One type involves the same scale $r_0'$ as in the equation of
motion [see Eqs. (\ref{2})-(\ref{4})]; the other one contains a
different length scale $r_0$, which is exactly the constant present in
the general formalism of Refs. \cite{BDI,B98tail,B98mult}. As we know
that the constant $r_0'$ is pure gauge, it will disappear from our
physical results at the end. As for $r_0$, it merely represents a
convenient scale entering the definition of the source multipole
moments in Ref. \cite{B98mult}, and should cancel out when considering
the complete multipole expansion of the field exterior to the
source. On the other hand, besides the harmless constants $r_0'$ and
$r_0$, there are three unknown dimensionless parameters in
Eq. (\ref{12}): $\xi$, $\kappa$ and $\zeta$. These parameters are
analogous to the constant $\lambda$ in the equations of motion (see
Ref. \cite{BIJ01} for their definition in the general case of
non-circular orbits). They probably reflect an incompleteness of the
standard Hadamard self-field regularization used in \cite{BIJ01}. It
is possible that the more sophisticated regularization proposed in
Ref. \cite{BFreg} could determine some (but maybe not all) of these
parameters. However, we shall see that, in the case of circular
orbits, the energy flux depends only on one combination of them:
$\theta=\xi+2\kappa+\zeta$, and furthermore that this constant
$\theta$ enters the energy flux at exactly the same level as
$\lambda$, so that the luminosity given by (\ref{9}) below depends on
one and only one combination of $\theta$ and $\lambda$ [to compute the
flux one needs the time derivatives of the moment (\ref{6}), and
$\lambda$ comes from replacing the accelerations by the equations of
motion (\ref{1})-(\ref{2})]. More work should be done to determine the
values of $\theta$ and $\lambda$.

Through 3.5PN order, the result concerning the ``instantaneous'' part of
the total energy flux, i.e. that part which is generated solely by the
multipole moments of the source (not counting the tails), is
\cite{BIJ01}
\begin{eqnarray}\label{7}
 {\cal L}_{\rm inst} &=& {32c^5\over 5G}\nu^2 \gamma^5 \biggl\{ 1 +
\left(-\frac{2927}{336}-\frac{5}{4}\nu \right) \gamma +
\left(\frac{293383}{9072}+\frac{380}{9}\nu\right) \gamma^2 \nonumber
\\ &+&\left(\frac{53712289}{1108800} -\frac{1712}{105}\ln
\left(\frac{r}{r_0}\right)\right.\nonumber\\&+&\left.
\left[-\frac{332051}{720}+\frac{123}{64}\pi^2+\frac{110}{3}\ln
\left(\frac{r}{r_0'}\right) +44\lambda-\frac{88}{3}\theta\right]\nu
-\frac{383}{9}\nu^2\right) \gamma^3\biggr\}\;,
\end{eqnarray}
where $\theta=\xi+2\kappa+\zeta$. The first term  represents the Newtonian
energy flux coming from the usual quadrupole formalism. To the latter
instantaneous part of the flux, we must add the non-linear tail
effects in the wave zone, which have already been calculated to the
3.5PN order in Ref. \cite{B98tail} (see Eqs. (5.5a) and (5.9)
there). We find
\begin{eqnarray}\label{8}
 {\cal L}_{\rm tail} &=& {32c^5\over 5G}\nu^2 \gamma^5 \biggl\{ 4\pi
\gamma^{3/2}+\left(-\frac{25663}{672}-\frac{125}{8}\nu\right)\pi
\gamma^{5/2}\nonumber\\
&+&\left(-\frac{116761}{3675}+\frac{16}{3}\pi^2-\frac{1712}{105}C
-\frac{856}{105}\ln (16~\!\gamma)+\frac{1712}{105}\ln
\left(\frac{r}{r_0}\right)\right) \gamma^3\nonumber\\ &+&
\left(\frac{90205}{576} +\frac{505747}{1512}\nu
+\frac{12809}{756}\nu^2\right)\pi \gamma^{7/2} \biggr\}\;,
\end{eqnarray}
where $C=.577\cdots$ denotes the Euler constant. What we call here
${\cal L}_{\rm tail}$ is in fact a complicated sum of ``tails'',
``tail squares'', and ``tails of tails'', as determined in
Ref. \cite{B98tail}. It is quite remarquable that so small an effect
as a ``tail of tail'', which constitutes the whole 3PN coefficient in
Eq. (\ref{8}), should be relevant to the present computation, which is
aimed at preparing the ground for a forthcoming experiment. As we can
see, the constant $r_0$ drops out from the sum of the instantaneous
(\ref{7}) and tail (\ref{8}) contributions --- which is normal, and
constitutes a first test of the calculation. However, the gauge
constant $r_0'$ does not seem to disappear at this stage, but that is
simply due to our use in Eqs. (\ref{7})-(\ref{8}) of the
post-Newtonian parameter $\gamma$, which depends {\it via} the
equation of motion on the choice of harmonic coordinates. After
substituting the frequency-related parameter $x$ in place of $\gamma$,
i.e. making consistent use of the relation (\ref{3}), we find that
$r_0'$ does cancel as well --- which represents another test, showing
the consistency between the two computations, in harmonic-coordinates,
of the equation of motion and multipole moments. Finally we obtain
\begin{eqnarray}\label{9}
 {\cal L} &=& {32c^5\over 5G}\nu^2 x^5 \biggl\{ 1 +
\left(-\frac{1247}{336}-\frac{35}{12}\nu \right) x + 4\pi
x^{3/2}\nonumber \\ &+&
\left(-\frac{44711}{9072}+\frac{9271}{504}\nu+\frac{65}{18}
\nu^2\right) x^2 +\left(-\frac{8191}{672}-\frac{583}{24}\nu\right)\pi
x^{5/2}\nonumber \\
&+&\left(\frac{6643739519}{69854400}+\frac{16}{3}\pi^2-\frac{1712}{105}C
-\frac{856}{105}\ln (16~\!x) \right.\nonumber\\
&+&\left.\left[-\frac{11497453}{272160}+\frac{41}{48}\pi^2
+\frac{176}{9}\lambda-\frac{88}{3}\theta\right]\nu-\frac{94403}{3024}\nu^2
-\frac{775}{324}\nu^3\right) x^3\nonumber\\ &+&
\left(-\frac{16285}{504}+\frac{214745}{1728}\nu
+\frac{193385}{3024}\nu^2\right)\pi x^{7/2} \biggr\}\;.
\end{eqnarray}
The last test (but not the least) is that the expression (\ref{9}) is
in perfect agreement, in the test-mass limit $\nu\to 0$ for one of the
bodies, with the result following from linear black-hole perturbations
obtained by Tagoshi and Sasaki \cite{TSasa94} (see also
Refs. \cite{Sasa94}). In particular, the rational fraction
$\frac{6643739519}{69854400}$, which is a sum of other fractions
appearing in both (\ref{7}) and (\ref{8}), comes out exactly the same
as in the black-hole perturbation theory \cite{TSasa94}.

We shall now deduce the laws of variation of the frequency and phase
using a balance equation as a fundamental tenet. Namely, we postulate
that
\begin{equation}\label{10}
{dE\over dt}=-{\cal L}\;,
\end{equation}
where the binary's gravitational binding energy $E$ is given by
Eq. (\ref{5}), and where the total gravitational-radiation luminosity
${\cal L}$ is the one obtained in Eq. (\ref{9}).  For justifications
of the validity of the energy balance equation (\ref{10}) in
post-Newtonian approximations, for either point-particle binaries or
extended weakly self-gravitating fluids, see
Refs. \cite{IW,B93}. Using the previous formulas for $E$ and ${\cal
L}$, we transform Eq. (\ref{10}) into an ordinary differential
equation for $\omega$, or, rather, the parameter $x$. For convenience
we adopt a new (dimensionless) time variable defined by
\begin{equation}\label{11}
\tau = \frac{\nu c^3}{5Gm}(t_c-t)\;,
\end{equation}
where $t_c$ denotes the instant of coalescence, at which the frequency
tends formally to infinity (evidently, the approximation breaks down
well before this point). The solution of the latter differential
equation reads
\begin{eqnarray}\label{12}
 x &=& {1\over 4}\tau^{-1/4}\biggl\{ 1 + \left( \frac{743}{4032}
 +\frac{11}{48}\nu\right)\tau^{-1/4} -
 \frac{1}{5}\pi\tau^{-3/8}\nonumber\\ &+& \left( \frac{19583}{254016}
 + \frac{24401}{193536} \nu + \frac{31}{288} \nu^2 \right) \tau^{-1/2}
 +\left(-\frac{11891}{53760} +\frac{109}{1920}\nu\right) \pi
 \tau^{-5/8}\nonumber\\
 &+&\left(-\frac{10052469856691}{6008596070400}+\frac{1}{6}\pi^2
 +\frac{107}{420}C
 -\frac{107}{3360}\ln\left(\frac{\tau}{256}\right)\right.\nonumber\\
 &+&\left.\left[\frac{15335597827}{3901685760}-\frac{451}{3072}\pi^2
 -\frac{77}{72}\lambda+\frac{11}{24}\theta\right]\nu
 -\frac{15211}{442368}\nu^2+\frac{25565}{331776}\nu^3\right)
 \tau^{-3/4}\nonumber\\ &+&
 \left(-\frac{113868647}{433520640}-\frac{31821}{143360}\nu
 +\frac{294941}{3870720}\nu^2\right)\pi \tau^{-7/8}\biggr\}\;.
\end{eqnarray}
Next we compute the binary's instantaneous phase, defined as the angle
$\phi$, oriented in the sense of the motion, between the separation of
the two bodies and, say, the direction of the ascending node of the
orbit within the plane of the sky. We have $\frac{d\phi}{dt}=\omega$
which translates, with our notation, into ${d\phi\over d\tau} =
-{5\over \nu} x^{3/2}$, and we can immediately integrate with the
result
\begin{eqnarray}\label{13}
\phi &=& - {1\over \nu} \biggl\{ \tau^{5/8} + \left( \frac{3715}{8064}
  + \frac{55}{96} \nu \right) \tau^{3/8} - \frac{3}{4}\pi\tau^{1/4}
  \nonumber\\ &+& \left( \frac{9275495}{14450688} +
  \frac{284875}{258048} \nu + \frac{1855}{2048} \nu^2 \right)
  \tau^{1/8} + \left( -\frac{38645}{172032} + \frac{65}{2048} \nu
  \right) \pi \ln \left(\frac{\tau}{\tau_0}\right)\nonumber\\
  &+&\left(\frac{831032450749357}{57682522275840}-\frac{53}{40}\pi^2
  -\frac{107}{56}C+\frac{107}{448}\ln\left(\frac{\tau}{256}\right)
  \right.\nonumber\\
  &+&\left.\left[-\frac{123292747421}{4161798144}+\frac{2255}{2048}\pi^2
  +\frac{385}{48}\lambda-\frac{55}{16}\theta\right]\nu
  +\frac{154565}{1835008}\nu^2-\frac{1179625}{1769472}\nu^3\right)
  \tau^{-1/8}\nonumber\\ &+&
  \left(\frac{188516689}{173408256}+\frac{488825}{516096}\nu
  -\frac{141769}{516096}\nu^2\right)\pi \tau^{-1/4}\biggr\}\;.
\end{eqnarray}
The constant $\tau_0$ is related to a constant phase that is simply
fixed by the initial conditions when the frequency of the wave enters
the detector's bandwidth. Finally it can be useful to dispose of the
expression of the phase in terms of the frequency $x$. For this we
have
\begin{eqnarray}\label{14}
\phi &=& - {1\over 32\nu}\biggl\{ x^{-5/2} + \left( \frac{3715}{1008}
 +\frac{55}{12}\nu\right)x^{-3/2} - 10\pi x^{-1}\nonumber\\ &+& \left(
 \frac{15293365}{1016064} + \frac{27145}{1008} \nu + \frac{3085}{144}
 \nu^2 \right) x^{-1/2} +\left(\frac{38645}{1344}
 -\frac{65}{16}\nu\right) \pi \ln
 \left(\frac{x}{x_0}\right)\nonumber\\
 &+&\left(\frac{12348611926451}{18776862720}-\frac{160}{3}\pi^2
 -\frac{1712}{21}C -\frac{856}{21}\ln (16~\!x)\right.\nonumber\\
 &+&\left.\left[-\frac{15335597827}{12192768}+\frac{2255}{48}\pi^2
 +\frac{3080}{9}\lambda-\frac{440}{3}\theta\right]\nu
 +\frac{76055}{6912}\nu^2-\frac{127825}{5184}\nu^3\right)
 x^{1/2}\nonumber\\ &+&
 \left(\frac{77096675}{2032128}+\frac{378515}{12096}\nu
 -\frac{74045}{6048}\nu^2\right)\pi x\biggr\}\;,
\end{eqnarray}
where $x_0$ is determined by the initial conditions (like
$\tau_0$). As a rough estimate of the relative importance of each of
the various post-Newtonian terms for LIGO/VIRGO-type detectors, we
give in Table I their contributions to the accumulated number $\cal N$
of gravitational-wave cycles (see also Table I in Ref. \cite{BDIWW95}
for the contributions of spin-orbit and spin-spin effects). 
\begin{table}
\caption{ Contributions to the accumulated number ${\cal N}={1\over
\pi}(\phi_{\rm ISCO}-\phi_{\rm seismic})$ of gravitational-wave
cycles. Frequency entering the bandwidth is $f_{\rm seismic}=$ 10 Hz;
terminal frequency is assumed to be at the Schwarzschild innermost
stable circular orbit $f_{\rm ISCO}=\frac{c^3}{6^{3/2}\pi G m}$. The
3PN term depends on the unknown parameter ${\hat \theta} = \theta -
\frac{7}{3}\lambda$ (we have ${\hat \theta} = \theta +
\frac{1987}{1320}$ using the value of $\lambda$ following from
$\omega_{\rm static}=0$).}
\begin{tabular}{lccc}
&$2 \times 1.4M_\odot$&$10M_\odot+1.4M_\odot$&$2 \times 10M_\odot$\\
\hline Newtonian &16031&3576&602\\ 1PN &441&213&59\\ 1.5PN
&$-$211&$-$181&$-$51\\ 2PN &9.9&9.8&4.1\\ 2.5PN
&$-$11.7&$-$20.0&$-$7.1\\ 3PN &2.5$+0.5~\!{\hat
\theta}$&2.2$+0.4~\!{\hat \theta}$&2.1$+0.4~\!{\hat \theta}$ \\ 3.5PN
&$-$0.9&$-$1.8&$-$0.8\\
\end{tabular}
\label{table1}
\end{table}
The result
for ${\cal N}_{\rm 3PN}$ is given as a function of the combination of
parameters ${\hat \theta} = \theta - \frac{7}{3}\lambda$ that enters
Eq. (\ref{14}). As we can see, {\it if} ${\hat \theta}$ is of the
order of unity, we reach with the 3PN or 3.5PN approximations an
acceptable level of, say, a few cycles, that roughly corresponds to
the demand which was made by data-analysists in the case of
neutron-star binaries \cite{3mn,CFPS}. Indeed, the above estimation
suggests that the neglected 4PN terms will yield some systematic
errors that are, at most, of the same order of magnitude, i.e. a few
cycles, and perhaps much less.  However, this conclusion is quite
sensitive to the exact value of ${\hat \theta}$. {\it If} ${\hat
\theta}$ is of the order of 10, we find that the 3PN term is nearly as
important numerically as the previous 2PN approximation. Finally, in
order to define the theoretical template of the compact binary
inspiral, one should insert the previous 3.5PN-accurate expressions of
the frequency and phase into the two polarisation wave-forms $h_+$ and
$h_\times$. A standard practice is to neglect in $h_+$ and $h_\times$
all the harmonics but the dominant one $f$ at twice the orbital
frequency (i.e. the so-called restricted post-Newtonian
approximation), but it is better to define the template by means of
the 2PN-accurate polarization wave-forms calculated in
Ref. \cite{BIWW96}.

G.F. acknowledges financial support of the EU-Network
HPRN-CT-2000-00137.


\end{document}